\documentclass{sigchi-ext}
\usepackage[T1]{fontenc}
\usepackage{textcomp}
\usepackage[scaled=.92]{helvet} 
\usepackage{graphicx} 
\usepackage{balance}  
\usepackage{booktabs} 
\usepackage{ccicons}  
\usepackage{ragged2e} 

\copyrightinfo{Permission to make digital or hard copies of all or part of this work for personal or classroom use is granted without fee provided that copies are not made or distributed for profit or commercial advantage and that copies bear this notice and the full citation on the first page. Copyrights for components of this work owned by others than the author(s) must be honored. Abstracting with credit is permitted. To copy otherwise, or republish, to post on servers or to redistribute to lists, requires prior specific permission and/or a fee. Request permissions from Permissions@acm.org.\\
{\emph{UbiComp/ISWC '16}} Adjunct, September 12-16, 2016, Heidelberg, Germany.\\
Copyright is held by the owner/author(s). Publication rights licensed to ACM. \\
ACM 978-1-4503-4462-3/16/09...\$15.00. \\
http://dx.doi.org/10.1145/2968219.2968544}

\def\plaintitle{Assessment of Social Roles for Interruption Management: A New Concept in the Field of Interruptibility} \def\plainauthor{Christoph Anderson, Clara Heissler, Sandra Ohly, Klaus David}

\def\plainkeywords{Interruptibility; Context-Aware Computing; Ubiquitous Computing; Social Roles.}

\title{Assessment of Social Roles for Interruption Management: A New Concept in the Field of Interruptibility}

\numberofauthors{4}
\author{%
  \alignauthor{%
    \textbf{Christoph Anderson}\\
    \affaddr{Wilhelmsh\"oher Allee 73} \\
    \affaddr{University of Kassel} \\
    \affaddr{34121 Kassel, Germany} \\
    \email{anderson@uni-kassel.de} }\alignauthor{%
    \textbf{Sandra Ohly}\\    
    \affaddr{Pfannkuchstra{\ss}e 1} \\
    \affaddr{University of Kassel} \\
    \affaddr{34121 Kassel, Germany} \\
    \email{ohly@uni-kassel.de} } \vfil \alignauthor{%
    \textbf{Clara Heissler}\\
    \affaddr{Pfannkuchstra{\ss}e 1} \\
    \affaddr{University of Kassel} \\
    \affaddr{34121 Kassel, Germany} \\
    \email{clara.heissler@uni-kassel.de} }\alignauthor{%
    \textbf{Klaus David}\\
    \affaddr{Wilhelmsh\"oher Allee 73} \\
    \affaddr{University of Kassel} \\
    \affaddr{34121 Kassel, Germany} \\
    \email{david@uni-kassel.de} \\
     } }

\definecolor{linkColor}{RGB}{6,125,233}
\hypersetup{%
  pdftitle={\plaintitle},
  pdfauthor={\plainauthor},
  pdfkeywords={\plainkeywords},
  bookmarksnumbered,
  pdfstartview={FitH},
  colorlinks,
  citecolor=black,
  filecolor=black,
  linkcolor=black,
  urlcolor=linkColor,
  breaklinks=true,
}

\begin{document}

\maketitle

\RaggedRight{} 

\begin{abstract}
Determining and identifying opportune moments for interruptions is a challenging task in Ubiquitous Computing and Human-Computer-Interaction. The current state-of-the-art approaches do this by identifying breakpoints either in user tasks, activities or by processing social relationships and contents of interruptions. However, from a psychological perspective, not all of these breakpoints represent opportune moments for interruptions. In this paper, we propose a \textit{new concept} in the field of interruptibility. The concept is based on role theory and psychological interruption research. In particular, we argue that \textit{social roles} which define sets of norms, expectations, rules and behaviours can provide useful information about the user's current context that can be used to enhance interruption management systems. Based on this concept, we propose a prototype system architecture that uses social roles to detect opportune moments for interruptions. 
\end{abstract}



\section{Introduction}
The question \textit{when} a system should interrupt their users has received much attention in the field of interruptibility. Most existing approaches determine opportune moments for interruptions by detecting breakpoints in tasks or activities. While these approaches consider only the external state (e.g. activity transition) of a user, not all of these breakpoints necessarily represent opportune moments for interruptions from a psychological point of view. The internal state of a user (e.g. intention) gives useful insights about disruptive interruptions caused by conflicting responsibilities such as work vs. family.

In fact, we hypothesize that the internal state of a user in combination with existing approaches needs to be considered in the field of interruptibility to reduce disruptive interruptions. To infer the internal state of a user, we leverage the concept of \textit{social roles}. According to role theory, \textit{social roles} are social positions that are linked to sets of norms, expectations, rules and behaviors a person is expected to comply with when enacting a role \cite{biddle_recent_1986}. Therefore, the current social role influences the actions of a user and determines the kind of information a user might prefer to receive or delay momentarily. However, roles typically belong to a specific life-domain and users might experience conflicts when role responsibilities from different domains are incompatible \cite{derks_2015}. 

The main contribution of this paper is to propose a \textit{new concept} for determining opportune moments for interruptions. By matching interruptions to a user's active role and therefore reducing disruptive interruptions that facilitate role conflicts, we hypothesize that interruption management systems can further be improved. In particular, (i) the current role of a user, (ii) the relevance of interruptions for this role, and (iii) the specific breakpoint that is chosen for an interruption should be considered in interruption management systems. These criteria have implications not only for the timing of interruptions but also for the information content a user receives.

\section{Opportune moments for interruptions}
Since interruptions are often unavoidable, it is important to time them in a way that adverse effects are reduced. A large body of existing research leverages the concept of breakpoints in \textit{tasks} and \textit{activities} to determine opportune moments for interruptions. In addition to these approaches, the \textit{content} of an interruption and the \textit{social relationship} between sender and recipient have also been examined. In general, approaches determine opportune moments by (i) identifying \textit{tasks} and \textit{task-boundaries}, (ii) processing \textit{activities}, \textit{device interactions} or \textit{application usage patterns} of users, (iii) utilizing \textit{content} and \textit{social relationship} between users. In the following sections, we give a short overview over existing approaches in these categories.

\subsection{Tasks and task boundaries}
A large body of research focuses on task-based models to predict opportune moments for interruptions. Most existing models are based on event perception techniques, which give insights about the decomposition of tasks in the human mind. In general, experiments with a predefined set of tasks are conducted. Participants are interrupted while working on tasks at a range of different time points. Effects on subjective workload \cite{adamczyk_2004}, mental and affective state \cite{pejovic_2015} as well as social attribution are measured \cite{iqbal_2005}.

Breakpoints between tasks are then used as opportune moments for interruptions. For example, Adamczyk et al. assume that breakpoints between coarser tasks should be chosen for interruptions as users are releasing cognitive resources after task completion that can be used for other peripheral tasks \cite{adamczyk_2004}. Another study focuses on task switching in multitasking environments rather than on the coarseness of breakpoints. The authors find that the number of interruptions, as well as duration, complexity and type of task influence the perceived difficulty of task switching and task resumption \cite{czerwinski_2004}. Their findings suggest that methods supporting the capturing and remembrance of suspended tasks may be helpful to assist users in multitasking environments.

\subsection{Activities, interactions and application usage patterns}
Also, \textit{physical activities} as well as \textit{application usage patterns} have been investigated as indicators for opportune moments. According to this approach, breakpoints in physical activities or application usage patterns define opportune moments for interruptions. Most of the reviewed research for this approach identifies breakpoints indicated by boundaries between two adjacent physical activities. Investigated activities for interruptibility encompass physical activities (e.g. sitting, standing, or walking \cite{ho_2005,okoshi_2015}), activities of daily living \cite{hudson_2003}, activities associated with semantic locations (e.g. \cite{sarker_2014}, as well as transportation types \cite{sarker_2014}. In addition to breakpoints in physical activities, the smartphone's posture and orientation \cite{poppinga_2014, turner_2015} have also been used to identify opportune moments for interruptions.

\begin{marginfigure}[-26pc]
  \begin{minipage}{\marginparwidth}
    \centering
    \includegraphics[width=0.9\marginparwidth]{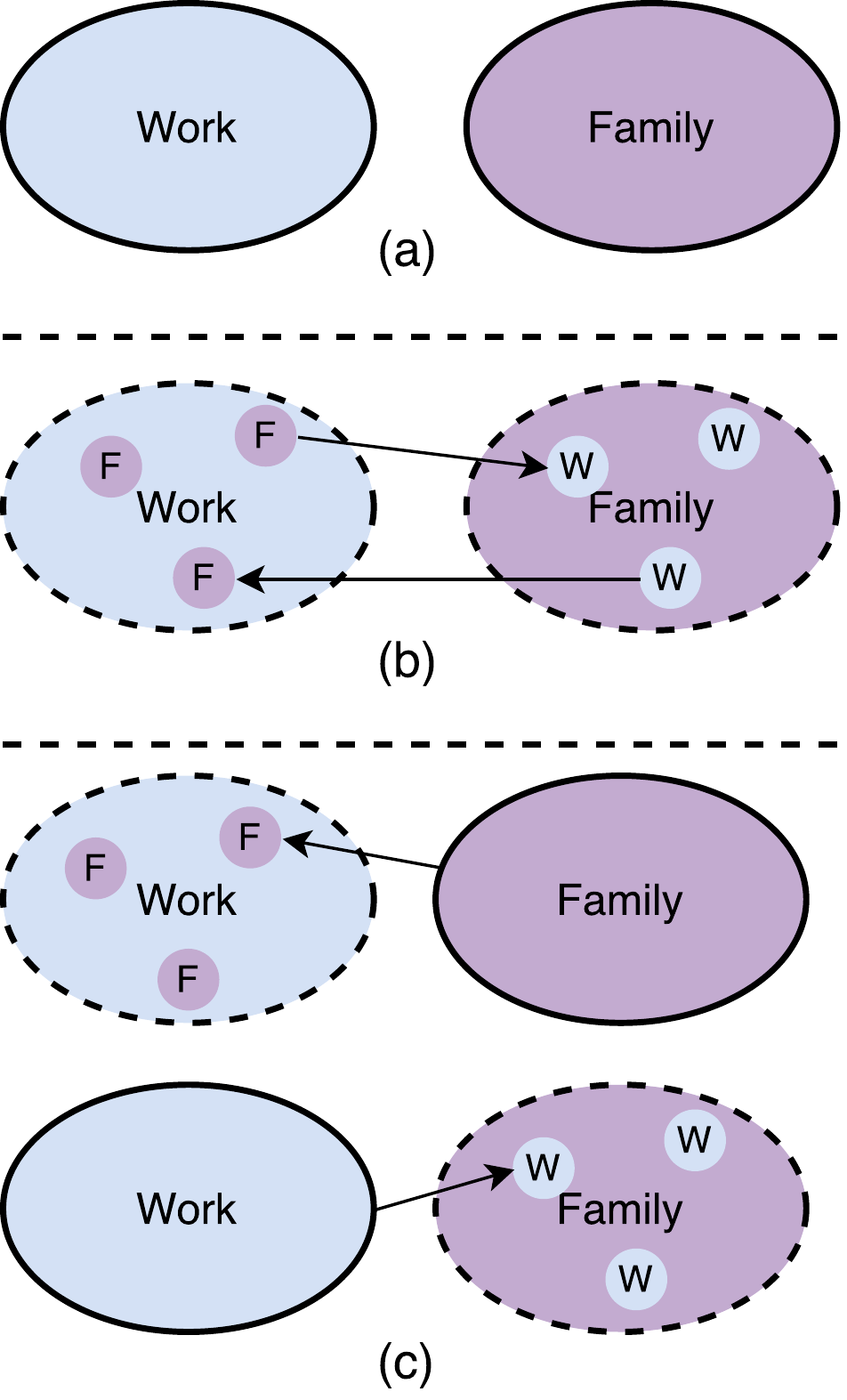}
    \caption{Different types of border preferences in a work-family scenario, a) segmentation, b) integration, c) combination}
    \label{fig:border_types}
  \end{minipage}
\end{marginfigure}

Another part of this research utilizes application usage patterns as an indicator for interruptibility. Similar to identifying breakpoints in physical activities, opportune moments for interruptions can be found by extracting breakpoints in application usage patterns (e.g. \cite{okoshi_2014, okoshi_2015b}). Recently, Okoshi et al. propose a breakpoint detection system combining physical activities and application usage patterns for reducing interruptive mobile notifications \cite{okoshi_2015b}. Breakpoints are identified by processing the event stream while users manipulated their smartphone. The authors utilize a multi-device approach including smartphones and smartwatches to determine breakpoints in physical activities as well as in application usage.

\subsection{Content and social relationship}
Another approach in the field of interruptibility determines opportune moments for interruptions by using the \textit{content} of interruptions (e.g. notification title on smartphones) as well as the \textit{social relationship} between sender and recipient. This approach utilizes information about the content and the initiator of the interruption to predict the interruptibility of users (e.g. \cite{fischer_2010, mehrotra_2015b}). Recently, Mehrotra et al. proposed an intelligent notification mechanism which combines both the content of an interruption and the social relationship between the sender and the recipient \cite{mehrotra_2015b}. They assume that social relationship can be assigned to one of four categories (work, social, family, or other) and define a category probability based on the current location of a user. Depending on the content of a notification title and the relationship between sender and recipient, the system predicts the interruptibility of users.

\section{Necessity of social roles for interruptibility}
Most of the approaches mentioned in the previous section consider the external state of a user such as the current activity or location to predict on the user's interruptibility. Some studies also investigate task engagement or importance of tasks to a user \cite{pejovic_2015}.

However, another aspect which gives insights about disruptive interruptions is the paradigm of \textit{cross-role interruptions} introduced by \textit{role theory}. The theory proposes that people adopt different roles with distinct norms, expectations, duties and rights in different life-domains. Also, people behave differently to adapt to specific demands and goals of a role. As roles typically belong to a specific life-domain, individuals might experience conflicts when role responsibilities from different domains are incompatible. In fact, the relation of smartphone usage and role conflicts has already been investigated in the field of psychology \cite{derks_2016,derks_2015}. It is well understood, that interruptions caused by conflicting social roles (e.g. calls from a supervisor while enacting a family role) result in interferences caused by incompatible norms and expectations. 

Nonetheless, the paradigm of cross-role interruptions has not been investigated in the field of interruptibility so far. For this reason, we propose to use information about the current social role of a user to recognize potential situations where role conflicts might occur.

\begin{marginfigure}[-6pc]
  \begin{minipage}{\marginparwidth}
    \includegraphics[width=0.9\marginparwidth]{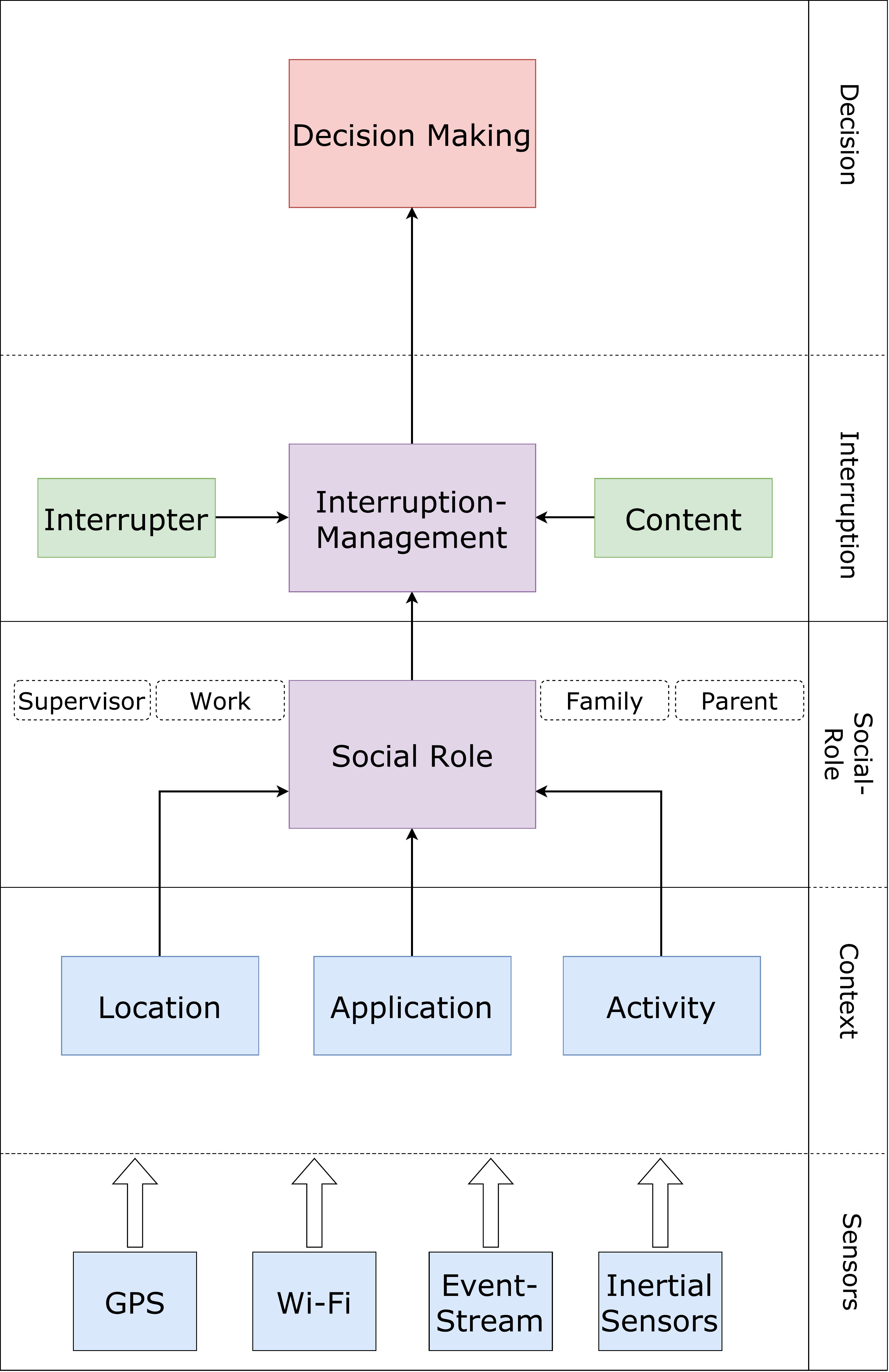}
    \caption{System architecture for inferring social roles of users}
    \label{fig:architecture}
  \end{minipage}
\end{marginfigure}

\section{System architecture to recognize social roles}
The current activity, ongoing tasks as well as communication with other people provide useful information about the current social role of a user. Therefore activity recognition along with information about principally used applications or visited locations is needed to assess social roles of users.

Figure \ref{fig:architecture} represents a first design concept of a system architecture that recognizes social roles of users. As already outlined in the previous section, social roles are defined by patterns of expectation, norms, and behaviors that apply to social positions. Therefore, our concept associates information about locations, activities, and applications of a user to predict social roles. We suggest that a user's behavior, expectation, and social role is reflected by (i) the type of application that is currently used, (ii) the location, (iii) and the current activity of a user. For example, if a user is interacting with an email client that is connected to his or her workplace, we can assume that this user might be working. In association with the current location and activity, a system can differentiate if the user is working at home or in a company, facilitating fine-grained assessments of social roles. By further processing information about the interrupter (e.g. social-relationship) and content of interruptions \cite{mehrotra_2015b}), an interruption management system can match all these information to detect potential cross-role interruptions. Individualized rules can then be applied to handle potential interruptions.

\section{Discussion}
Currently, a prototype system that is based on the proposed concept is being built. A diary study is also being conducted to verify our approach of using social roles in the field of interruptibility. Nonetheless, further investigation of the utilization of social roles is needed. 

In particular, different preferences how social roles and life-domains are separated and structured need to be examined for such an interruption management system. For example, people who prefer \textit{segmentation} use strong and non-permeable boundaries to separate different life-domains (e.g. work vs. family). Erecting strong boundaries means that \textit{segmenters} try to avoid cross-role interruptions. They might be more negatively affected by those interruptions than people who integrate or use a combination of segmentation and integration tactics (see Figure \ref{fig:border_types}).

\textit{More precisely, how can a system automatically detect different border preferences of users?}
These and other open research questions need to be examined in the future when determining opportune moments for interruptions by using social roles. 

\section{Conclusion}
In this paper, we present a new concept for predicting the interruptibility of users based on role theory and psychological interruption research. In particular, our concept is based on \textit{social roles} of users which reflect sets of norms, expectations, rules and behaviours a person is expected to comply with when enacting a role. We hypothesize, that when \textit{interruptions} are matched to \textit{social roles} of users, interruptions are likely to be more relevant for users and less disruptive. Therefore, we assume that a user's \textit{current role} is a key aspect that needs to be considered and investigated for further research in the field of interruptibility.

\section{Acknowledgements}
This work has been funded by the Social Link Project within the Loewe Program of Excellence in Research, Hessen, Germany
\balance{}

\bibliographystyle{SIGCHI-Reference-Format}
\bibliography{bibliography}


\begin{thebibliography}{00}


\ifx \showCODEN    \undefined \def \showCODEN     #1{\unskip}     \fi
\ifx \showDOI      \undefined \def \showDOI       #1{{\tt DOI:}\penalty0{#1}\ }
  \fi
\ifx \showISBNx    \undefined \def \showISBNx     #1{\unskip}     \fi
\ifx \showISBNxiii \undefined \def \showISBNxiii  #1{\unskip}     \fi
\ifx \showISSN     \undefined \def \showISSN      #1{\unskip}     \fi
\ifx \showLCCN     \undefined \def \showLCCN      #1{\unskip}     \fi
\ifx \shownote     \undefined \def \shownote      #1{#1}          \fi
\ifx \showarticletitle \undefined \def \showarticletitle #1{#1}   \fi
\ifx \showURL      \undefined \def \showURL       #1{#1}          \fi

\bibitem{adamczyk_2004}
{P.~D. Adamczyk} {and} {B.~P. Bailey}. 2004.
\newblock \showarticletitle{If {Not} {Now}, when?: {The} {Effects} of
  {Interruption} at {Different} {Moments} {Within} {Task} {Execution}}. In {\em
  Proceedings of the {SIGCHI} {Conference} on {Human} {Factors} in {Computing}
  {Systems}} {\em ({CHI} '04)}. ACM, New York, NY, USA, 271--278.
\newblock
\showISBNx{1-58113-702-8}
\showDOI{%
\url{http://dx.doi.org/10.1145/985692.985727}}


\bibitem{biddle_recent_1986}
{B.~J. Biddle}. 1986.
\newblock \showarticletitle{Recent {Developments} in {Role} {Theory}}.
\newblock {\em Annual Review of Sociology\/} {12}, 1 (1986), 67--92.
\newblock
\showDOI{%
\url{http://dx.doi.org/10.1146/annurev.so.12.080186.000435}}


\bibitem{czerwinski_2004}
{M. Czerwinski}, {E. Horvitz}, {and} {S. Wilhite}. 2004.
\newblock \showarticletitle{A {Diary} {Study} of {Task} {Switching} and
  {Interruptions}}. In {\em Proceedings of the {SIGCHI} {Conference} on {Human}
  {Factors} in {Computing} {Systems}} {\em ({CHI} '04)}. ACM, New York, NY,
  USA, 175--182.
\newblock
\showISBNx{1-58113-702-8}
\showDOI{%
\url{http://dx.doi.org/10.1145/985692.985715}}


\bibitem{derks_2016}
{D. Derks}, {A.~B. Bakker}, {P. Peters}, {and} {P. van Wingerden}. 2016.
\newblock \showarticletitle{Work-related smartphone use, work--family conflict
  and family role performance: The role of segmentation preference}.
\newblock {\em Human Relations\/} {69}, 5 (2016), 1045--1068.
\newblock
\showDOI{%
\url{http://dx.doi.org/10.1177/0018726715601890}}


\bibitem{derks_2015}
{D. Derks}, {D. van Duin}, {M. Tims}, {and} {A.~B. Bakker}. 2015.
\newblock \showarticletitle{Smartphone use and work--home interference: The
  moderating role of social norms and employee work engagement}.
\newblock {\em Journal of Occupational and Organizational Psychology\/} {88}, 1
  (2015), 155--177.
\newblock
\showISSN{2044-8325}
\showDOI{%
\url{http://dx.doi.org/10.1111/joop.12083}}


\bibitem{fischer_2010}
{J.~E. Fischer}, {N. Yee}, {V. Bellotti}, {N. Good}, {S. Benford}, {and} {C.
  Greenhalgh}. 2010.
\newblock \showarticletitle{Effects of {Content} and {Time} of {Delivery} on
  {Receptivity} to {Mobile} {Interruptions}}. In {\em Proceedings of the 12th
  {International} {Conference} on {Human} {Computer} {Interaction} with
  {Mobile} {Devices} and {Services}} {\em ({MobileHCI} '10)}. ACM, New York,
  NY, USA, 103--112.
\newblock
\showISBNx{978-1-60558-835-3}
\showDOI{%
\url{http://dx.doi.org/10.1145/1851600.1851620}}


\bibitem{ho_2005}
{J. Ho} {and} {S.~S. Intille}. 2005.
\newblock \showarticletitle{Using {Context}-aware {Computing} to {Reduce} the
  {Perceived} {Burden} of {Interruptions} from {Mobile} {Devices}}. In {\em
  Proceedings of the {SIGCHI} {Conference} on {Human} {Factors} in {Computing}
  {Systems}} {\em ({CHI} '05)}. ACM, New York, NY, USA, 909--918.
\newblock
\showISBNx{1-58113-998-5}
\showDOI{%
\url{http://dx.doi.org/10.1145/1054972.1055100}}


\bibitem{hudson_2003}
{S. Hudson}, {J. Fogarty}, {C. Atkeson}, {D. Avrahami}, {J. Forlizzi}, {S.
  Kiesler}, {J. Lee}, {and} {J. Yang}. 2003.
\newblock \showarticletitle{Predicting Human Interruptibility with Sensors: A
  Wizard of Oz Feasibility Study}. In {\em Proceedings of the SIGCHI Conference
  on Human Factors in Computing Systems} {\em (CHI '03)}. ACM, New York, NY,
  USA, 257--264.
\newblock
\showISBNx{1-58113-630-7}
\showDOI{%
\url{http://dx.doi.org/10.1145/642611.642657}}


\bibitem{iqbal_2005}
{S.~T. Iqbal} {and} {B.~P. Bailey}. 2005.
\newblock \showarticletitle{Investigating the {Effectiveness} of {Mental}
  {Workload} {As} a {Predictor} of {Opportune} {Moments} for {Interruption}}.
  In {\em {CHI} '05 {Extended} {Abstracts} on {Human} {Factors} in {Computing}
  {Systems}} {\em ({CHI} {EA} '05)}. ACM, New York, NY, USA, 1489--1492.
\newblock
\showISBNx{1-59593-002-7}
\showDOI{%
\url{http://dx.doi.org/10.1145/1056808.1056948}}


\bibitem{mehrotra_2015b}
{A. Mehrotra}, {M. Musolesi}, {R. Hendley}, {and} {V. Pejovic}. 2015.
\newblock \showarticletitle{Designing {Content}-driven {Intelligent}
  {Notification} {Mechanisms} for {Mobile} {Applications}}. In {\em Proceedings
  of the 2015 {ACM} {International} {Joint} {Conference} on {Pervasive} and
  {Ubiquitous} {Computing}} {\em ({UbiComp} '15)}. ACM, New York, NY, USA,
  813--824.
\newblock
\showISBNx{978-1-4503-3574-4}
\showDOI{%
\url{http://dx.doi.org/10.1145/2750858.2807544}}


\bibitem{okoshi_2014}
{T. Okoshi}, {J. Nakazawa}, {and} {H. Tokuda}. 2014.
\newblock \showarticletitle{Attelia: {Sensing} {User}'s {Attention} {Status} on
  {Smart} {Phones}}. In {\em Proceedings of the 2014 {ACM} {International}
  {Joint} {Conference} on {Pervasive} and {Ubiquitous} {Computing}: {Adjunct}
  {Publication}} {\em ({UbiComp} '14 {Adjunct})}. ACM, New York, NY, USA,
  139--142.
\newblock
\showISBNx{978-1-4503-3047-3}
\showDOI{%
\url{http://dx.doi.org/10.1145/2638728.2638802}}


\bibitem{okoshi_2015b}
{T. Okoshi}, {J. Ramos}, {H. Nozaki}, {J. Nakazawa}, {A.~K. Dey}, {and} {H.
  Tokuda}. 2015a.
\newblock \showarticletitle{Attelia: {Reducing} user's cognitive load due to
  interruptive notifications on smart phones}. In {\em 2015 {IEEE}
  {International} {Conference} on {Pervasive} {Computing} and {Communications}
  ({PerCom})}. IEEE, St. Louis, MO, USA, 96--104.
\newblock
\showDOI{%
\url{http://dx.doi.org/10.1109/PERCOM.2015.7146515}}


\bibitem{okoshi_2015}
{T. Okoshi}, {J. Ramos}, {H. Nozaki}, {J. Nakazawa}, {A.~K. Dey}, {and} {H.
  Tokuda}. 2015b.
\newblock \showarticletitle{Reducing {Users}' {Perceived} {Mental} {Effort}
  {Due} to {Interruptive} {Notifications} in {Multi}-device {Mobile}
  {Environments}}. In {\em Proceedings of the 2015 {ACM} {International}
  {Joint} {Conference} on {Pervasive} and {Ubiquitous} {Computing}} {\em
  ({UbiComp} '15)}. ACM, New York, NY, USA, 475--486.
\newblock
\showISBNx{978-1-4503-3574-4}
\showDOI{%
\url{http://dx.doi.org/10.1145/2750858.2807517}}


\bibitem{pejovic_2015}
{V. Pejovic}, {M. Musolesi}, {and} {A. Mehrotra}. 2015.
\newblock \showarticletitle{Investigating {The} {Role} of {Task} {Engagement}
  in {Mobile} {Interruptibility}}. In {\em Proceedings of the 17th
  {International} {Conference} on {Human}-{Computer} {Interaction} with
  {Mobile} {Devices} and {Services} {Adjunct}} {\em ({MobileHCI} '15)}. ACM,
  New York, NY, USA, 1100--1105.
\newblock
\showISBNx{978-1-4503-3653-6}
\showDOI{%
\url{http://dx.doi.org/10.1145/2786567.2794336}}


\bibitem{poppinga_2014}
{B. Poppinga}, {W. Heuten}, {and} {S. Boll}. 2014.
\newblock \showarticletitle{Sensor-{Based} {Identification} of {Opportune}
  {Moments} for {Triggering} {Notifications}}.
\newblock {\em IEEE Pervasive Computing\/} {13}, 1 (Jan. 2014), 22--29.
\newblock
\showISSN{1536-1268}
\showDOI{%
\url{http://dx.doi.org/10.1109/MPRV.2014.15}}


\bibitem{sarker_2014}
{H. Sarker}, {M. Sharmin}, {A.~A. Ali}, {Md.~M. Rahman}, {R. Bari}, {S.~M.
  Hossain}, {and} {S. Kumar}. 2014.
\newblock \showarticletitle{Assessing the {Availability} of {Users} to {Engage}
  in {Just}-in-time {Intervention} in the {Natural} {Environment}}. In {\em
  Proceedings of the 2014 {ACM} {International} {Joint} {Conference} on
  {Pervasive} and {Ubiquitous} {Computing}} {\em ({UbiComp} '14)}. ACM, New
  York, NY, USA, 909--920.
\newblock
\showISBNx{978-1-4503-2968-2}
\showDOI{%
\url{http://dx.doi.org/10.1145/2632048.2636082}}


\bibitem{turner_2015}
{L.~D. Turner}, {S.~M. Allen}, {and} {R.~M. Whitaker}. 2015.
\newblock \showarticletitle{Push or {Delay}? {Decomposing} {Smartphone}
  {Notification} {Response} {Behaviour}}.
\newblock In {\em Human {Behavior} {Understanding}}, {Albert~Ali Salah}, {Ben
  J.~A. Kr{\"o}se}, {and} {Diane~J. Cook} (Eds.). Number 9277 in Lecture
  {Notes} in {Computer} {Science}. Springer International Publishing, Cham,
  Switzerland, 69--83.
\newblock
\showISBNx{978-3-319-24194-4 978-3-319-24195-1}
\showDOI{%
\url{http://dx.doi.org/10.1007/978-3-319-24195-1_6}}


\end{thebibliography}

\end{document}